\documentclass[12pt]{article}
\usepackage{amsmath}
\usepackage{graphicx,psfrag,epsf}
\usepackage{enumerate}
\usepackage{natbib}
\usepackage{amsmath,amsthm,amsfonts,epsfig,amssymb,natbib,eucal,eufrak}
\usepackage{booktabs}
\usepackage{multirow}
\usepackage{siunitx}
\usepackage{qtree}
\usepackage{color}

\usepackage{float}
\floatstyle{plaintop}
\restylefloat{table}

\newtheorem{result}{Result}

\newcommand{\blind}{0}

\begin{document}

\def\spacingset#1{\renewcommand{\baselinestretch}%
{#1}\small\normalsize} \spacingset{1}


\if0\blind
{
  \title{\bf Revisiting nested group testing procedures: new results, comparisons and robustness}
  \author{Yaakov Malinovsky\thanks{Corresponding author}\\
    Department of Mathematics and Statistics\\ University of Maryland, Baltimore County, Baltimore, MD 21250, USA\\
    and \\
    Paul S. Albert\thanks{
    The work was supported by the National Cancer Institute Intramural Program.}\hspace{.2cm}\\
    Biostatistics Branch, Division of Cancer Epidemiology and Genetics\\National Cancer Institute, Rockville, MD 20850, USA}

  \maketitle
} \fi

\if1\blind
{
  \bigskip
  \bigskip
  \bigskip
  \begin{center}
    {\LARGE\bf Title}
\end{center}
  \medskip
} \fi

\bigskip
\begin{abstract}
Group testing has its origin in the identification of syphilis in the  US army during World War II. Much of the theoretical framework of group testing
was developed starting in the late 1950s, with continued work into the 1990s. Recently, with the advent of new laboratory and genetic technologies, there has been
an increasing interest in group testing designs for cost saving purposes.
In this paper, we compare different nested designs, including Dorfman, Sterrett and an optimal nested procedure obtained through dynamic programming.
To elucidate these comparisons, we develop closed-form expressions for the optimal Sterrett procedure and provide a concise review of the prior literature for other commonly used procedures. We consider designs where the prevalence of disease is known as well as investigate the robustness of these procedures when it is incorrectly assumed.
This article provides a technical presentation that
will be of interest to researchers as well as from a pedagogical perspective.
Supplementary material for this article is available online.
\end{abstract}

\noindent%
{\it Keywords:} Coding theory; information theory; optimal design.
\vfill

\newpage
\spacingset{1.45} 
\section{Introduction}
\label{se:I}
Commonly, individual samples are assessed for the presence of a condition in order to identify disease status. Group testing is concerned with finding
efficient algorithms to test groups of individual samples that provide these identifications with a minimum number of tests.
Group testing (GT) procedures are cost- and time-saving identification procedures that have broad applications to blood screening for HIV, hepatitis and other infectious diseases (\cite{G1994}; \cite{BT2010}; \cite{S2011}; \cite{T2013}; \cite{B2016}),
quality control in product testing (\cite{SG1959}; \cite{BP1990}), veterinary medicine (\cite{G2016}), drug discovery (\cite{Z2001}), DNA screening
(\cite{Dh2006}; \cite{CS2016}), communication and security networks (\cite{W1985}; \cite{L12013}),
and experimental physics (\cite{BG2000}; \cite{MBR2009}), among others.

Although group testing has its roots in the
complete identification of a given population with respect to a particular disease,
the estimation of parameters from probability models has also been considered.
For example, there is an extensive literature on group testing for disease prevalence estimation
(\cite{To1962}; \cite{Tu1995}; \cite{DH2012}; \cite{Liu2012}; \cite{Wa2016}; \cite{Ha2016}). This article deals only with the identification problem.

There are two main classifications of group testing models for identification purposes:
probabilistic group testing (PGT) and combinatorial group testing (CGT).
In PGT, a probability model is assumed for the joint distribution of $N$ binary population items.
In CGT, it is assumed that there are a certain number of infected individuals among $N$ individuals
and combinatorial techniques are used to identify them (\cite{Dh1999}).
For a more detailed discussion of this classification, see \cite{B2005}.
This article focuses on only PGT.

Throughout this article, we assume that the tests are not subject to misclassification (i.e., a gold-standard test is assumed).
In many practical situations, a test is only considered for screening when the misclassification is very small.
Thus, it is important to present a careful comparison of designs when tests
are not subject to error.
However, we do recognize the work of others who have addressed the issue of screening with a misclassified test \citep{Kim2007,M2012,MAR2016}.

The purpose of this article is to provide a theoretical framework for comparing commonly used designs for group testing.
In order to rigourously perform these comparisons, we develop new and discuss previously known results in the group testing (GT) literature.
An early PGT formulation was a simple procedure proposed by Dorfman \citep{Dorfman1943}, followed by an extension proposed by Sterrett \citep{S1957}.
Although these two procedures have been investigated, no one has presented a careful comparison of these two designs relative to an optimal nested procedure that
we will define later.

We start the discussion with three commonly used GT procedures that were designed for the binomial group testing problem, where a set of $N$ individuals have to be classified either as positive or not under assumptions
that each individual has the same probability $p$ of being positive and the outcomes of different individuals are independent. In the group testing literature such a set is called the {\it binomial set} \citep{SG1959}.
\\

\noindent
{\it Dorfman procedures: Procedures $D$} and $D^{'}$\\
We begin by introducing the Dorfman blood testing problem (see also \cite{F1950}, p. 189).
The motivation was the need to administer blood tests for syphilis to millions of people drafted into the US army during World War II.
\cite{Dorfman1943} suggested to group the blood samples of each of the $k$ people and apply a single blood test to the entire group.
If the group test is negative, then only the single test is required for identification of the $k$ individuals.
If the group test is positive then each of the $k$ individuals is tested separately, resulting in $k+1$ tests.
This procedure is commonly referred to as the Dorfman two-stage group testing procedure (Procedure $D$).
The intuition behind Procedure $D$ is that for small $p$, a second stage will rarely be required.
This method is used by the American Red Cross in the screening of blood donations for HIV and hepatitis \citep{D2002}.

Dorfman's two-stage GT procedure belongs to the nested class of GT procedures, which we will define later. The performance of any GT procedure will be evaluated with respect to the expected number of tests. We call a GT procedure {\it optimal} within a particular  class of procedures if it has the minimal expected number of tests.

There is a logical inconsistency in Procedure {\it D}.
It is clear that any ``reasonable" group testing plan should satisfy the following property:
``A test is not performed if its outcome can be inferred from previous test results" (\cite{U1960}, p. 50).
Procedure $D$ does not satisfy this property
since if the group is positive and all but the last person are negative, the last person is still tested.
The modified Dorfman procedure \citep{SG1959} (defined as $D^\prime$) would not test the last individual in this case.

Even though the two procedures $D$ and $D^{\prime}$ are very similar, we will show later in this article that
particularly, when the prevalence is high,
the efficiency gain of $D^{\prime}$ over $D$ with respect to the expected number of tests may be substantial.
\\

\noindent
{\it Sterrett procedure: Procedure $S$}\\
\cite{S1957} suggested an improvement of Procedure $D^\prime$ in the following way.
If in the first stage of Procedure $D^\prime$ the group is positive, then in the second stage
individuals are tested one-by-one until the first positive individual is identified.
Then the first stage of Procedure $D^\prime$ is applied to the remaining (nonidentified) individuals.
The procedure is repeated until all individuals are identified.

When $p$ is small, the probability of having two or more positive individuals in a group is very small.
Therefore, when retesting a positive group, it is most probable that we will test sequentially until the first positive
and then the remaining individuals grouped together will be negative. Thus, it is intuitively clear that Procedure $S$ will be more efficient
than Procedure $D^\prime$ in this situation.
\\

\noindent
{\it Efficiency of a GT procedure}\\
When comparing the different procedures throughout this article,
we make the distinction between an infinite and a finite population of size $N$.
This is done since for a large (infinite) population it is natural to obtain optimality results under the assumption of equal group sizes, and for a finite population
we need to consider situations where the population cannot be divided into equal group sizes.
More formally, for an infinite population we are looking for a common group size $k^{*}_{A}$ (for a given Procedure $A$ $\left(A\in (D, D^{\prime}, S)\right)$) that minimizes the expected number of tests per person.
For a finite $N$, we partition the population into subsets where we apply a Procedure $A$ $\left(A\in (D, D^{\prime}, S)\right)$
within each subset. In this case, the optimality is defined by finding a partition such that the expected total number of tests is minimal.
For example, when $N=10$, we may consider a design where we apply the procedure to the entire population, partition the 10 individuals into two groups of size
5, or partition them into three groups of size $3, 3$ and $4$, and apply the procedure separately in each group.
In Section \ref{sec:Finite}, we will show that the optimal partition has equal subgroup sizes if $N$ is divisible by $k^{*}_{A}$.
This motivates us to start the discussion  with the infinite population case in order to obtain $k^{*}_{A}$,
and then to apply the optimality results from the infinite to the finite population case.

For a general group testing problem, a procedure is optimal if, for a given $N$ and $p$, it achieves the minimum expected total number of tests $E(N,p)$.
A general optimal procedure is unknown, and its characterization was conjectured as an intractable problem \citep{DK1987}.
\cite{DK1987} determined the computational complexity of a wide class of group testing models, where
they proved that a general version of binomial group testing is NP-complete (no polynomial time solution is known) \citep{GJ1979}.

There are only a few fundamental results in binomial group testing that provide insights on the structure of $E(N,p)$.
\cite{U1960} characterized the optimality of any group testing algorithm and proved that if $p\geq p_{U}=(3-5^{1/2})/2\approx 0.38$, then there does not exist an algorithm that is better than individual one-by-one testing. That is,
$$E(N,p)=N,\,\,\,\,\text{for}\,\,\,\,\,p\geq p_{U}.$$
Web Appendix G provides additional details about Ungar's result. From now on, we will refer to $p_{U}$ as Ungar's universal cut-off point (UCP). Another important result is due to \cite{YH1988}, who showed that $E(N,p)$ is increasing in $p$ for $0<p< p_{U}$ and $N\geq 2$.

A nested algorithm has the property that if the positive subset $I$ is identified, the next subset $I_1$ that we will test is a proper subset of $I$, that is, $I_1\subset I$. This natural class of GT procedures was defined by \cite{SG1959} and \cite{S1960}, and it is clear that Procedures $D$, $D^{\prime}$ and $S$ all belong to this class.
The optimal nested algorithm is not optimal among all possible GT algorithms, but they are simple to implement due to their sequential nature. Further, the optimal nested algorithm
is nearly optimal over all algorithms \citep{S1960, S1967}.

This article is organized as follows. In order to compare procedures, we present in Section \ref{sse:BC} the expected number of tests per person under Procedures $D$, $D^\prime$ and $S$. A new short proof for the expected number of tests under Procedure $S$ is obtained.
In Sections \ref{sse:DInf}, \ref{sse:DprimeInf} and \ref{sse:Sinf},  we present the optimization problem under Procedures $D$, $D^\prime$ and $S$ for the infinite population, and
in Section \ref{sec:Finite}, we present an optimal partition of the finite population under Procedures $D$, $D^{\prime}$ and $S$.
In Section \ref{sec:DP}, we present the optimal nested procedure that can be found (even for a very large population size) with dynamic programming.
Section \ref{sec:r} investigates the robustness of an optimum nested procedure versus $D$, $D^{\prime}$ and $S$ in the case where the only available information is an upper bound on parameter $p$.
The proofs of key optimality results are presented in Appendices A to E. Web Appendices F to I provide theoretical derivations to support other results stated in this article. Web Appendix J provides Matlab code for the optimum nested procedure.

\section{Optimality under an infinite population}
\label{sec:Inf}
In this section, we will compare two simple procedures that are useful for the case of (large) infinite populations.
In order to make these comparisons, we need to present some known as well as new theoretical results.

\subsection{Expected number of tests for the Procedures $D$, $D^{\prime}$ and $S$}
\label{sse:BC}
We denote the expected number of tests per person in a group of size $k$ under Procedure $A$ as  $E_{A}\left(k,\, p\right)$.
Under the binomial group testing model, we have the following characteristics under Procedures $D$, $D^{\prime}$ and $S$.\\
\\
\noindent
{\it Procedure $D$}\\
For $k\geq 2,$ the total number of tests is $1$ with probability $q^k$ ($q=1-p$) and $k+1$ with probability $1-q^k$. Therefore,
\begin{equation}
\label{eq:D}
E_{D}\left(k, p\right)=
\left\{\begin{array}{ccc}
                          \displaystyle   1-q^{k}+\frac{1}{k} & for & k\geq 2 \\
                            1 & for & k=1.
                          \end{array}
                          \right.
\end{equation}

\noindent
{\it Procedure $D^\prime$}\\
For $k\geq 2,$  the total number of tests is $1$ with probability $q^k$, $k$ with probability $q^{k-1}(1-q)$,
and $k+1$ with probability $1-q^k-q^{k-1}(1-q)=1-q^{k-1}$. Therefore,
\begin{equation}
\label{eq:D'}
E_{{D^\prime}}\left(k, p\right)=
                            \displaystyle 1-q^k+1/k-(1/k)(1-q)q^{k-1}.
\end{equation}
It is easy to check that $\displaystyle E_{D^{\prime}}\left(1, p\right)=1$, and that
$E_{{D^\prime}}\left(k, p\right)\leq E_{{D}}\left(k, p\right)$.

\medskip
\noindent
{\it Procedure S}\\
\cite{S1957} provided an expression for $\displaystyle E_{S}\left(k, p\right)$ in terms of a finite sum of terms
for which each element involves a binomial coefficient, and the resulting expression is very complex.
In fact, this finite sum has a simple closed-form expression.
\begin{result}
\label{res:1}
\begin{equation}
\label{eq:S}
\displaystyle
E_{S}\left(k, p\right)=   \frac{1}{k}\left[2k-(k-2)q-\frac{1-q^{k+1}}{1-q}\right].
\end{equation}
\end{result}
\noindent
It is easy to check that $\displaystyle E_{S}\left(1, p\right)=1$.
\cite{SG1959} provided this closed-form expression for $\displaystyle E_{S}\left(k, p\right)$
as a consequence of the general recursive equations.
We prove this result with alternative short arguments in Appendix A.

In the remainder of this section, we present results needed for a careful comparison of Procedures $D, D^{\prime}$ and $S$.
\subsection{Determining the optimal design for procedures $D, D^{\prime}$ and $S$ }
\label{sec:LPC}
For an infinite population, our goal is to find the optimal group size $k^{*}_{A}(p)$ for a given Procedure $A$.
It should be recognized that $k^{*}_{A}(p)$ is a function of $p$ and this dependence on $p$ is suppressed in the notation.
The difficulty in developing a closed-form expression for  $k^{*}_{A}$ lies in the discreteness of the problem.
Also, equations \eqref{eq:D} and \eqref{eq:D'} are not unimodal as a function of $k$ for a given $p$.

\subsubsection{Procedure $D$}
\label{sse:DInf}
In the original work of \cite{Dorfman1943}, there is no closed-form solution for $k^{*}_{D}(p)$, only numerical evaluations.
For Procedure $D$, \cite{S1978} solved this optimization problem and showed that if $p<p_{D}=1-1/3^{1/3}\approx 0.31,$ then
$k^{*}_{D}$ is equal to $\displaystyle 1+[p^{-1/2}]$ or $\displaystyle 2+[p^{-1/2}]$ (where [p] is denoted as the integer part of $p$); otherwise, $\displaystyle k^{*}_{D}=1$.
From his result, it follows that
the applicability of Procedure $D$ is limited by the value of $p_{D}$.

\subsubsection{Procedure $D^\prime$}
\label{sse:DprimeInf}
Procedure $D^\prime$ was mentioned by \cite{SG1959} and investigated in detail by \cite{PE1978}. They did not provide the closed-form solution for $k^{*}_{D^{\prime}}$ but provided the following result, which immediately led to the solution.
\\
\\
\noindent
{\it Lemma 2 in \cite{PE1978}}.\\
{\it Let $\displaystyle p\in \left(0, (3-\sqrt{5})/2\right)$. Then (as a function of the continuous variable $k$) $\displaystyle E_{D^{\prime}}\left(k, p\right)$ has an absolute minimum which is at the smallest zero of $\displaystyle E^{\,\,'}_{D^{\prime}}\left(k, p\right)=\frac{\partial E_{D^{\prime}}\left(k, p\right)}{\partial k}$.
This zero is unique in that portion of the domain of  $\displaystyle E_{D^{\prime}}$ for which $\displaystyle E_{D^{\prime}}\left(k, p\right)<1$.}
\\
\noindent
From the above lemma, it follows that the optimal value $k^{*}_{D^{\prime}}$ is the smallest $k$ value which satisfies
\begin{align}
\label{eq:PE}
&
E_{D^{\prime}}\left(k, p\right)\leq E_{D^{\prime}}\left(k-1, p\right)\,\,\,\,\text{and}\,\,\,\,E_{D^{\prime}}\left(k, p\right)< E_{D^{\prime}}\left(k+1, p\right).
\end{align}
Therefore, we have to sequentially check the above inequalities for $k=2,3,\ldots$ in order to find this smallest value of $k$.

It is clear from the above inequalities that there are nonunique solutions for some values of $p$ (i.e., there are two solutions for the value of $p$ where $\displaystyle E_{D^{\prime}}\left(k, p\right)= E_{D^{\prime}}\left(k-1, p\right)$).
From equations \eqref{eq:D} and \eqref{eq:D'} and Lemma 2 of \cite {PE1978}, it follows that  $\displaystyle k^{*}_{D^{\prime}}\leq k^{*}_{D}$
for $\displaystyle p<1-1/3^{1/3}.$
It was stated in \cite{PE1978} that it does not
seem possible to explicitly obtain a closed-form expression for the optimal group size $k^{*}_{D^{\prime}}(p)$.
Although we cannot prove it, we empirically verified the conjecture that the optimal group size $k^{*}_{D^{\prime}}(p)$ is
equal to $\displaystyle \lfloor p^{-1/2}\rfloor$ or $\displaystyle \lceil p^{-1/2}\rceil$ for $ 0<p<p_U=\frac{3-\sqrt{5}}{2}$,
where $\displaystyle \lfloor x\rfloor \left(\displaystyle \lceil x \rceil\right)$ for $x>0$ is defined as the largest (smallest) integer which is smaller (larger) or equal to $x$.
This conjecture was examined for values of $p$ in the above range with incremental steps of $10^{-6}$ in the following way. For a given value of $p$,
the optimal group size $k^{*}_{D^{\prime}}(p)$ was found using  \eqref{eq:PE}, and it was then verified that it is equal to either $\displaystyle \lfloor p^{-1/2}\rfloor$ or $\displaystyle \lceil p^{-1/2}\rceil$.
The curious reader can easily check this conjecture  empirically using the values of Table \ref{t:1} and Table 1 in \cite{PE1978}.


\begin{table}[H]
\caption{
The minimal (optimal) expected number of tests per 100 individuals $\left(100E_{A}\left(k^*_{A}, p\right)\right)$ for Procedure $A$ $\left(A\in\left\{D,D^{\prime},S\right\}
\right)$ using an optimal group size $k^*_{A}$ for different $p$
}
\small
\begin{center}
  \begin{tabular}{lllllll}
    \toprule
    \multirow{2}{*}{p} &
      \multicolumn{2}{c}{$D$ } &
      \multicolumn{2}{c}{$D^{\prime}$ } &
      \multicolumn{2}{c}{$S$} \\
      & {$k^*_{D}$} & {$100E_{D}\left(k^*_{D}, p\right)$} & {$k^*_{D^{\prime}}$} & {$100E_{D^{\prime}}\left(k^*_{D^{\prime}}, p\right)$} & {$k^*_{S}$} & {$100E_{S}\left(k^*_{S}, p\right)$} \\
      \midrule
    0.001 & 32    & 6.2759  &         32 &      6.2729 &          45 &       4.5844 \\
    0.005 &15 &       13.91&           15&       13.879&           21&       10.535\\
    0.01 &11      & 19.557&           10&        19.47&           15&       15.172\\
    0.03 & 6  &     33.369&            6&        32.94&            9&       27.305\\
    0.05 &  5 &      42.622&            5  &     41.807&            7&       35.977\\
    0.07 & 4     &  50.195&            4&       48.787&            6&       43.167\\
    0.10  &4 &       59.39&            4&       57.567&            5&       52.288\\
    0.13 &3&       67.483&            3&       64.203&            4&       60.042\\
    0.15&3       &71.921&            3&       68.308&            4&       64.784\\
    0.20 & 3       &82.133&            3&       77.867&            3&       74.933\\
    0.25& 3      & 91.146     &       2 &      84.375 &           3 &      83.854\\
    0.27& 3    &   94.432&            2&       86.855&            2&       86.855\\
    0.30&  3     &  99.033&            2&         90.5&            2&         90.5\\
    0.32& 1      &100 &           2&        92.88&            2&        92.88\\
    0.35&  1      &100&            2&       96.375&            2&       96.375\\
    0.38&  1     &100 &           2&        99.78&            2&        99.78\\
   \bottomrule
  \end{tabular}
  \label{t:1}
  \end{center}
  \end{table}

\subsubsection{Procedure $S$}
\label{sse:Sinf}
\cite{S1957} failed to provide the closed-form expression for $E_{S}\left(k, p\right)$ but instead provided a large-sample (infinite population) approximation.
As a consequence, there are some inaccurate results in his  Table I.

The following new result provides a way to find the optimal group size under Procedure $S$.
\begin{result}
\label{res:2}
Let $\displaystyle p\in \left(0, (3-\sqrt{5})/2\right)$. Then (as a function of continuous variable $k$, $k\geq 1)$ $\displaystyle E_{S}\left(k, p\right)$ has an absolute minimum at the unique zero of $\displaystyle E^{\,\,'}_{S}\left(k, p\right)$.
\end{result}

For the proof of Result \ref{res:2}, please see Appendix B.\\
From Result \ref{res:2}, it follows that the optimal value $k^{*}_{S}(p)$ is equal to $\displaystyle \lceil l \rceil$ or $ \displaystyle \lfloor l \rfloor$, where $\displaystyle E^{\,\,'}_{S}\left(l, p\right)=0.$  Alternatively, to avoid solving the nonlinear equation $E^{\,\,'}_{S}\left(l, p\right)=0$, we can find the optimal $k^{*}_{S}$ in the same manner as under Procedure $D^{\prime}$ using \eqref{eq:PE}.
We conjecture that the optimal group size $k^{*}_{S}(p)$ is
equal to $\displaystyle \lfloor \sqrt{2/p}\rfloor$ or $\displaystyle \lfloor \sqrt{2/p}\rfloor +1$ or $\displaystyle \lfloor \sqrt{2/p}\rfloor +2$ for $ 0<p<p_U$.
Although we cannot prove it, we empirically verified this conjecture in the same manner as we did for $k^{*}_{D^{\prime}}(p)$ in Section \ref{sse:DInf}.

\subsubsection{A comparison of Procedures $D$, $D^{\prime}$, and $S$ }
Table \ref{t:1} shows a detailed comparison of the optimal group size and the corresponding total expected number of tests per 100 for Procedures $D$, $D^{\prime}$ and $S$ as a function of $p$. In particular, for large $p$, Procedure $D^{\prime}$ has an impressive efficiency gain over Procedure $D$. Although Procedure $S$ is uniformly better than $D^{\prime}$ for all $p$, for small values of $p$, Procedure $S$ is substantially better than $D^{\prime}$. \cite{S1957} provided a similar comparison
for Procedures $D$ and $S$. However, he only used  large sample approximations for the optimal group size and expected number of tests, which were slightly  inaccurate.
Table \ref{t:1} along with previous discussed theoretical results show that
Procedures $D^{\prime}$ and $S$ (but not $D$) achieve the same upper applicability bound (UCP) $p_{U}$
(see also Web Appendix H).

\section{Optimality under a finite population}
\label{sec:Finite}
In Section \ref{sec:Inf}, we have discussed the infinite population case, where for a given Procedure $A\in \left\{D, D^{\prime}, S\right\}$,  we find the value $k$ which minimizes the expected number of tests per person, $\displaystyle E_{A}\left(k,\, p\right)=\frac{h_{A}(k,\, p)}{k}$, where $\displaystyle h_{A}(k,\,p)$ is the expected total number of tests for a group of size $k$ and for a prevalence of $p$.
Define $\displaystyle h_{A}(k)=h_{A}(k,\,p)$.

Generally, for a finite population of size $N$ and a given Procedure $A$, we have to solve the following optimization problem:
find the optimal {\it partition}
$\displaystyle \left\{n_1,\ldots,n_I\right\}$ with $n_1+\ldots+n_{I}=N$ for some $I\in\left\{1,\ldots,N\right\} $ such that $E_{A}\left(k,\, p\right)$  is minimal (denote $H_{A}(N)$), that is, $\displaystyle \left\{n_1,\ldots,n_I\right\}$ is a solution of the following optimization problem:
\begin{equation}
\label{eq:DP}
\displaystyle
H_{A}\left(N\right)=\min_{m_1,\,m_2,\ldots,m_J}\sum_{i=1}^{J}h_{A}(m_i),\,\,\text{subject to},\,\,\,\sum_{i=1}^{J}m_i=N,\,\,\,J\in\left\{1,\ldots,N\right\}.
\end{equation}

Recall from the Introduction that for $p\geq p_U$, the optimum is to test one-by-one for either finite or infinite population size.
Therefore, in this case $I=N$ and the optimal partition is $\left\{n_i=1,\,\,i=1,\ldots,N\right\}.$
A common method to solve \eqref{eq:DP} is dynamic programming (DP) \citep{B1957}; the first application of DP in group testing appeared in \cite{SG1959}, and that for Procedure $D^{\prime}$ was presented by \cite{PE1978}.
The DP algorithm can be expressed as
\begin{align}
\label{eq:DPS}
&h_{A}(1)=1,\,\,\,H_{A}(0)=0,\,\,\,H_{A}(1)=1,\nonumber \\
&
H_{A}(k)=\min_{0\leq x \leq k-1} \left\{H_{A}\left(x\right)+h_{A}(k-x)\right\},\,\,\,\,\,k=2,\ldots,N.
\end{align}

It is obvious that the computation effort of the above DP algorithm is $O(N^2)$, which makes it easy to implement, and it can be computationally fast enough for even large values of $N$.

We can compare $D$, $D^{\prime}$ and $S$ for a finite population using this DP algorithm \eqref{eq:DPS}.
For example, if $p=0.05$, then the optimal group size for an infinite population under Procedure $D$ is $k^{*}_{D}=5$ (Table 1),
and the optimal partition when $N=13$  is $\displaystyle \left\{n_1,n_2,n_3\right\}=\left\{5,4,4\right\}$ with $\displaystyle H_{D}\left(13\right)=5.615$;
for Procedure $D^{\prime}$ is $k^{*}_{D^{\prime}}=5$ (Table 1), and the optimal partition when $N=13$  is $\displaystyle \left\{n_1,n_2,n_3\right\}=\left\{5,4,4\right\}$ with $\displaystyle H_{D^{\prime}}\left(13\right)=5.489$;  for Procedure $S$ with the same $p=0.05$,  $k^{*}_{S}=7$ (Table 1) and the optimal partition when $N=13$ is $\displaystyle\left\{n_1,n_2\right\}=\left\{6,7\right\}$ with $\displaystyle H_{S}\left(13\right)= 4.685$.

The example above  illustrates some interesting features of subgroup sizes of the optimal partition. Specifically, we see that the optimal partition subgroup sizes differ at most by one unit. This was conjectured by \cite{LS1972} as a general result for Procedure $D$ and they provided insight by using the convexity (with respect to $k$) property of an approximation to  $E_{D}\left(k,\, p\right)$. \cite{G1985} proved this result for Procedure $D^{\prime}$. We will prove this result for Procedure $S$.

In the following result, we show that the optimal partition has equal subgroup sizes if $N$ is divisible by $k^{*}_{S}$.
It is clear from the proof that the same holds for the Procedures $D$ and $D^{\prime}$.
\begin{result}
\label{res:d}
Suppose we apply the group testing algorithm $S$ for a finite population of size $N$ for a given $p$.
Also suppose that $N=sk^{*}_{S}(p)$, i.e. $s$ subgroups of size $k^{*}_{S}$. Then the optimal partition is $\left\{n_i=k^{*}_{S}, i=1,\ldots,s\right\}$, i.e., $I=s$ and the infinite population optimal solution is the subgroup size of the optimal partition for the finite population.
\end{result}

\begin{proof}
Please see Appendix C.
\end{proof}

The following result establishes a relationship among the size of subgroups under the optimal partition for Procedure $S$.
\begin{result}
\label{res:LS}
Suppose we apply group testing algorithm $S$ for a finite population of size $N$ for a given $p$ ($p \in (0,1)$) and we start with some partition
$\displaystyle \left\{m_1,\ldots,m_J\right\}$. There exists a better (with respect to expected number of tests) partition $\left\{m^{'}_1,\ldots,m^{'}_J\right\}$  with $\displaystyle |m^{'}_j-m^{'}_i|\leq 1$ for all $i,j$.
\end{result}

\begin{proof}
The proof is based on the convexity property of $h_{S}(x)$ with respect to $x$ (see also discussion on the previous page) and is presented in Appendix D.
\end{proof}

From Result \ref{res:LS}, it follows that if we apply group testing algorithm $S$ for a finite population of size $N$ for a given $p$ ($p \in (0,1)$) and we start with an optimal number of subgroups $I$, then there exists an optimal partition containing groups whose sizes differ by at most 1.
\\
\noindent

The following result, which was conjectured for Procedures $D$ and $D^{\prime}$ by \cite{LS1972}, provides a simple way to construct an optimal partition.
\cite{G1985} proved the result for $D^{\prime}$ and his method also applies to Procedure $D$. We prove it for Procedure $S$.

\begin{result}
\label{res:Main}
Suppose we apply group testing algorithm $S$ for a finite population of size $N$ for a given $p$ ($p \in (0,1)$).
Denote $a$ to be an optimal group size under Procedure $S$ for an infinite population, $s=\lfloor\frac{N}{a}\rfloor$ (i.e., $s$ groups of size $a$) and $\theta=N-sa$ (i.e., remainder $0< \theta <a$).
Then the optimal partition is one of the two following partitions:
\begin{itemize}
\item[(i)]
Distribute the remainder $\theta$ among $s$ groups (with initial size $a$) in such a way that $|n_i-n_j|\leq 1$ for all $i,j \in \left\{1,\ldots,s\right\}$.
\item[(ii)]
Build up an additional group (group $s+1$) by taking the remainder $\theta$ and units from the above $s$ groups (with initial size $a$) in such way that $|n_i-n_j|\leq 1$ for all $i,j \in \left\{1,\ldots,s,s+1\right\}$.
\end{itemize}
\end{result}

\begin{proof}
Please see Appendix E.
\end{proof}

From Result \ref{res:Main}, it follows that in order to find the optimal partition, we need to evaluate the total expected number of the tests in (i) and (ii)
and choose the design ((i) or (ii)) that minimizes this quantity.
Result \ref{res:Main} provides for finding the optimum partition without any computational cost. In Web Appendix F (Remark 1), we provide an alternative direct implementation of Result \ref{res:Main}.

Table \ref{t:2} provides the optimum partition (Procedures $D$, $D^{\prime}, S$) for the finite population case for different values of $p$ in a similar way as in Table 1.
The last two columns of Table \ref{t:2} will be discussed in Section \ref{sec:NPr}.

\begin{table}[H]
\caption{
The minimal (optimal) expected number of tests per 100 individuals $\left(H_{A}(100)\right)$ for Procedure A $ \left(A\in\left\{D, D^\prime, S\right\}\right)$, and Procedure $\,R_1$ $\left(E_{1}(100)\right)$ and comparison with the information lower bound $H(p)$. OP means an optimal partition, and $s\times a$ means $s$ groups of size $a$   }
\footnotesize
\label{t:2}
\begin{center}
  \begin{tabular}{lllllll|llllllll}
    \toprule
    \multirow{2}{*}{p} &
      \multicolumn{2}{c}{D } &
      \multicolumn{2}{c}{$D^{\prime}$ } &
      \multicolumn{2}{c}{$S$} \\
    & {OP} & {$H_{D}(100)$} & {OP} & {$H_{D^{\prime}}(100)$} & {OP} & {$H_{S}(100)$} &{$E_{1}(100)$}& $H(p)$ \\
      \midrule
    0.001 & $2\times 33, 1\times 34$ &6.281&  $2\times 33, 1\times 34$ &    6.278 &$2\times 50$ &                4.605 & 1.766&  1.141\\
    0.005 &$5\times 14, 2\times 15$ & 13.917&          $5\times 14, 2\times 15$ &        13.884&           $5\times 20$&       10.537 &4.749&4.541\\
    0.01 &$10\times 10$ &19.562&          $10\times 10$ &        19.470& $5\times 14, 2\times 15$&       15.181 &  8.320& 8.079\\
    0.03 & $12\times 6, 4\times 7$ &33.402&          $12\times 6, 4\times 7$ &        32.993&$10\times 9, 1\times 10$&       27.325  &19.693&19.439 \\
    0.05 &$20\times 5$   &42.622&          $20\times 5$  &     41.807&$5\times 6, 10\times 7$&       36.018& 28.958&28.640 \\
    0.07&$25\times 4$ & 50.195&           $25\times 4$&       48.787&$2\times 5, 15\times 6$&       43.184&  36.916& 36.592\\
    0.10  &$25\times 4$ &59.390&           $25\times 4$&       57.567&$20\times 5$&       52.288& 47.375&  46.900\\
    0.13 &$32\times 3, 1\times 4$ &67.492&           $32\times 3, 1\times 4$&       64.258&$25\times 4$&       60.042 &56.183 &  55.744\\
    0.15&$32\times 3, 1\times 4$  &71.956&          $32\times 3, 1\times 4$&       68.396&$25\times 4$&       64.784& 61.485 &  60.984\\
    0.20 &$32\times 3, 1\times 4$ &82.210&           $2\times 2, 32\times 3$&       77.872&$32\times 3, 1\times 4$&       74.974& 72.875& 72.192\\
    0.25&$32\times 3, 1\times 4$ &91.234&       $50\times 2$ &      84.375 &$2\times 2, 32\times 3$&      83.875&  82.191&  81.128\\
    0.27&$32\times 3, 1\times 4$ &94.518&           $50\times 2$&       86.855&$50\times 2$&86.855&84.864&84.146\\
    0.30&$32\times 3, 1\times 4$ &99.117&           $50\times 2$&         90.5&$50\times 2$&90.5&88.889&88.129\\
    0.32&$100\times 1$ &100&       $50\times 2$&        92.88&$50\times 2$&92.88&91.574&90.438\\
    0.35&$100\times 1$ &100&         $50\times 2$&       96.375&$50\times 2$&96.375&95.633&93.407\\
    0.38&$100\times 1$ &100&       $50\times 2$&        99.78&$50\times 2$&99.78&99.730&95.804\\
   \bottomrule
  \end{tabular}
  \end{center}
  \end{table}

\section{Optimal nested procedure and connection with coding theory }
\label{sec:NPr}
\subsection{Optimal nested procedure}
A nested procedure, which was defined in the Introduction \citep{SG1959}, requires that between any two successive tests:
\begin{itemize}
\item[(i)]
future tests are concerned only with units not yet classified as good or defective,
\item[(ii)]
$n$ units not yet classified have to be separated into only (at most) two sets. One set of size $m\geq 0$, called the ``defective set," is known to contain at least one defective unit if $m\geq 1$ (it is not known which ones are defective or exactly how many there are). The other set of size $n-m\geq 0$ is called the ``binomial set" because we have no knowledge about it other than the original binomial assumption.
Either of these two sets can be empty in the course of experimentation; both are empty at termination.
\end{itemize}

\label{sec:DP}
The number of potential nested group testing algorithms is astronomical. For example, if $N=5$, then there are  $235,200$ possible algorithms \citep{MS1977}.
Therefore, it is impossible to directly evaluate the expected number of tests for each algorithm, making a direct computation infeasible.
\cite{SG1959} overcame this problem by proposing a DP algorithm that
finds the optimal nested algorithm, which Sobel and Groll termed ``the Procedure $R_1$."
There was a large research effort to reduce the computational complexity $O(N^3)$ of the original proposed algorithm \citep{S1960,KS1971,H1976}.
With new theoretical results, \cite{S1960}
reduced the complexity to $O(N^2)$. Further, \cite{KS1971} reduced the computation complexity by half as compared with \cite{S1960}. Finally,
\cite{H1976}, using the results for optimal binary trees (Huffman trees, \cite{H1952}) and optimal alphabetic binary trees \citep{HT1971}, reduced the computational complexity to $O(N)$ (not including the sorting effort).
In addition, \cite{YH1990} proved that
the pairwise testing algorithm (groups of size two) is the unique (up to the substitution of equivalent items) optimal nested algorithm for all $N$ if and only if
$\displaystyle 1-\frac{1}{\sqrt{2}}<p<p_{U}$ (at the boundary values the pairwise testing algorithm is an optimal nested algorithm). Recently, \cite{ZP2016} provided
an asymptotic analysis of the optimal nested procedure.
\\
\noindent
The development of the optimal nested algorithm (due to Sobel) of complexity $O(N^2)$ is presented in Web Appendix I. This result allows for computing the optimal total expected number of tests $H_1(N)$ (under the optimal nested Procedure $R_1$).
\\
\noindent
For example, if $p=0.05$ and $N=13$, then the expected number of tests under the optimum nested Procedure $R_1$ is $H_1(13)=3.878$. For comparison, with the same
values of $p$ and $N$, we obtain $\displaystyle H_{D}(13)=5.615$, $\displaystyle H_{D^{\prime}}(13)=5.489$ and $\displaystyle H_{S}(13)=4.685$.
Web Appendix I provides a thorough explanation of the construction of an optimal nested procedure in this case.

In Table \ref{t:2}, we present the expected number of tests (optimum nested procedure) per $N=100$ individuals ($E_{1}(100)$)  for different values of $p$.

\subsection{Coding theory and Information lower bound}
In the previous subsection, we showed that DP can be used to obtain the optimal nested procedure.
However, it does not speak more generally to optimality among all possible procedures.
Deriving an information lower bound for the expected total number of tests of an optimal procedure provides insight into the efficiency of the optimal nested procedure. The information lower bound (ILB) was provided in \cite{SG1959}. \cite{S1960, S1967} use noiseless-coding theory to derive ILB .
In Web Appendix H, we carefully demonstrate the development of the ILB using coding theory attributed to \cite{S1960, S1967}.
Web Appendix H provides a good pedagogical tool for this important development in GT.
The key result is the information lower bound $H(p)$ for the expected number of tests under an optimal procedure that is the Shannon formula of entropy:
\begin{align}
&
H(p)=N \left[p\log_{2}\frac{1}{p}+q\log_{2}\frac{1}{q}\right].
\end{align}
The information lower bound $H(p)$ is not attainable but provides a benchmark for what is a close-to-attainable level for an optimal group testing procedure
(for a detailed discussion, see Web Appendix H).
In the last column of Table \ref{t:2}, we present the information lower bound $H(p)$ for different values of $p$ when $N=100$.

\section{Robustness investigation}
\label{sec:r}
In this section, we investigate the robustness of the procedures to the incorrect specification of the parameter $p$.
In order to simplify this investigation, we will assume the large population setting, which will allow for a common group size for a given procedure.
Optimal group sizes under Procedures $D$, $D^{\prime}$, $S$, and an optimal nested procedure are all functions of parameter $p$. However, $p$ may not be known, and interest is on the comparison of different design strategies when $p$ is not correctly specified.
In some situations, there is only knowledge of an upper bound $U$ of the design parameter $p$. Under a constant group size setting, such as in Procedures $D$, $D^{\prime}$ and $S$, we can follow the methodology developed by \cite{MA2015} to calculate the minimax group size $k^{**}_A$ for Procedures $A \in \left\{D, D^{\prime}, S\right\}$ as
\begin{equation}
\label{eq:mm}
k^{**}_{A}= \arg\min_{k \in \mathbb{N}^{+}}\sup_{p\in (0, U]} L_{A}\left(k, p\right),
\end{equation}
where $\displaystyle L_{A}\left(k, p\right)=E_{A}\left(k, p\right)-E_{A}\left(k^{*}(p), p\right),\,\,\,A \in \left\{D, D^{\prime}, S\right\}.$

Table \ref{table:4} shows the expected number of tests per 100 individuals for minimax designs of $D, D^{\prime}$ and $S$ along with nested Procedure $R_1$ using
$U$ instead of $p$ ($H_1(100)$) and $U/2$ instead of $p$ ($H_1^{*}(100)$) for different $p$.
We evaluated the nested Procedure $R_1$ at a value of $U$. Further,
since values of $p$ are often substantially lower than a specified upper bound $U$, we also evaluated the procedure at a value of $U/2$.

\begin{table}[!h]
\scriptsize
\caption{Robustness of the nested Procedure $R_1$ vs Procedures $D$, $D^{\prime}$ and $S$} 
\begin{center}
 \begin{tabular}{ l|cccc|cccc|ccccc}
 \hline
\multicolumn{1}{c}{}&\multicolumn{4}{ c| }{$U=0.05$}&\multicolumn{4}{ c| }{$U=0.10$}&\multicolumn{4}{ c }{$U=0.20$} \\
\hline
 $p$                             & $0.001$ & $0.005$ & $0.01$&$0.05$ & $0.001$ & $0.01$ &$0.05$ &$0.10$ & $0.001$  &$0.01$  & $0.1$ &$0.2$ \\
   \hline
 $100E_{D}\left(k^{**}_{D}, p\right)$   &$10.185$&$14.455$&$19.557$&$52.211$&$13.297$&$20.226$&$46.158$&$69.453$&$13.297$&$20.226$&$69.453$&$95.723$\\
  \hline
 $100E_{D^{\prime}}\left(k^{**}_{D^{\prime}}, p\right)$  & $   10.985$ &$14.841$ & $ 19.470$ & $ 49.811$&$13.285$ & $ 20.109$ & $45.721$ & $68.855$&$ 14.969$ & $ 20.945$ & $65.697$ &$92.565$\\
 \hline
 $100E_{S}\left(k^{**}_{S}, p\right)$   & $ 7.975$ &$ 11.241$ & $ 15.185$ &$41.899$& $10.628$ & $16.138$ & $ 37.760$ &  $59.381$&$ 13.024$ & $ 17.647$ & $ 55.928$ & $85.889$\\
 \hline
 $H_{1}(100)$                      & $7.468$ &$9.311$ & $11.567$ & $ 28.958$&$15.287$ & $18.007$ & $30.979$ &$ 47.375$&  $33.233$ & $  35.221$ & $ 53.271$ & $72.875$\\
 \hline
 $H_{1}^{*}(100)$                      & $4.511$ &$6.578$ & $ 9.194$ & $30.242$&$7.468$ & $11.567$ & $ 28.958$ & $50.282$& $15.287$ & $   18.007$ & $ 47.375$ & $79.988$\\
 \hline
 $k^{**}_{D}$ &$11$&$11$&$11$&$11$&$8$&$8$&$8$&$8$&$8$&$8$&$8$&$8$&\\
 \hline
 $k^{**}_{D^{\prime}}$                   &$10$&$10$&$10$&$10$&$8$&$8$&$8$&$8$&$7$&$7$&$7$& $7$\\
 \hline
 $k^{**}_{S}$                   &$14$&$14$&$14$&$14$&$10$&$10$&$10$&$10$&$8$&$8$&$8$& $8$\\
 \hline
 \end{tabular}
\label{table:4} 
\end{center}
\end{table}

Table \ref{table:4} shows that Procedure $R_1$ is generally more efficient than $D$, $D^{\prime}$ and $S$. However, in rare situations where the assumed upper bound $(U)$ is substantially
higher than the true unknown $p$, Procedure $S$ may indeed be more efficient than $R_1$.

\section{Summary}
This article provides a unique perspective on group testing where we tie together literature on infinite and finite populations GT,
dynamic programming and coding theory.
This is done in order to
compare important nested group testing procedures, including Dorfman ($D$ and $D^{\prime}$), Sterrett (S)
and an optimal nested Procedure
$R_1$, with the theoretical information lower bound of efficiency serving as a reference.
All theoretical developments were essential for making these comparisons.

Some of the results were provided previously in the literature, while others, particulary
for Procedure $S$, are new. We demonstrated that, particularly when $p$ is small, Procedure $S$ has a large efficiency gain relative to Procedures
$D$ and $D^{\prime}$.
Further, there can be a sizable efficiency gain by using the optimal nested procedure that is based on DP. However, this efficiency gain needs to be weighed against
the practical complications in implementing the different procedures. For example, although there is a sizable efficiency gain in using the optimal nested procedure,
the complex nature of the design may make it less practical (see Web Appendix I).
The less efficient Dorfman procedure is simple to implement in that it is a two-stage procedure, where
testing within the second stage can be conducted in parallel (simultaneously).
This is in contrast with the Sterrett procedure,
where stages subsequent to the first stage are sequential and cannot be performed in parallel.

These results are based on the correct specification of $p$. Using our newly derived results on the Sterrett procedure, we were able to show that even when $p$ is misspecified,
the optimal nested procedure is generally more efficient than $S$. However, it is important to recognize that this may not be the case when our knowledge of $p$ is far from the truth (i.e., when $U$ is substantially larger than $p$). Importantly, for any $p$, Procedure $S$ is more efficient than $D^{\prime}$.

The results in this paper highlight the importance of studying efficient procedures in group testing. Simplicity aside, the Sterrett and optimal nested procedures
are more efficient than Dorfman's procedure. This is also true when $p$ is misspecified. Modern applications of nested procedures have generally focused on using
Dorfman's procedure (\cite{Hill2016}; \cite{F2015}),
although Sterrett's procedure is also used (\cite{BT2010}),
rather than an alternative nested procedure.
The results of this article clearly demonstrate the advantage of using the Sterrett procedure or an optimal nested procedure whenever it practicably feasible. Based on our results, we encourage the use of the Sterrett procedure and optimal nested procedures in practice.

\section*{Acknowledgments}
The authors thank the editor, the associate editor and two referees for their thoughtful and constructive comments.
Also, the comments and suggestions of the associate editor resulted in significant improvements of the article.
The authors thank Mattson Publishing Services for editing the article.
This paper is dedicated to the memory of David Assaf, a teacher and a friend.

\bigskip





\section*{Appendix}
\appendix
\section{Proof of Result 1}
\label{A:a}
\begin{proof}
Let $X$ be the number of tests in order to identify $k$ persons and let $1_j$ be an indicator function that is equal to 1 if the first positive identified person is the person $j$ ($j=1,\ldots,k$) tested.
Also denote $1_0$ as an indicator function that is equal to $1$ if no positive person is in the group of size $k$.
We have
$$X=X1_0+X1_1+X1_2+\ldots+X1_{k-1}+X1_k.$$
Define $\displaystyle E_{k}(X)\equiv E(k)=kE_{S}\left(k,p\right)$.  It is clear that $\displaystyle E(1)=1$.
Therefore,
\begin{align*}
&E(k)=q^k+kq^{k-1}(1-q)+(1-q)\left(2+E(k-1)\right)\\
&
+q(1-q)\left(3+E(k-2)\right)+q^2(1-q)\left(4+E(k-3)\right)+\ldots+q^{k-2}(1-q)\left(k+E(k-(k-1))\right)\\
&
=
1-(k-1)q^k+\frac{1-q^{k-1}}{1-q}+(1-q)\left[E(k-1)+qE(k-2)+q^2E(k-3)+\ldots+q^{k-2}E(1)\right].
\end{align*}
Taking the difference $\displaystyle E(k+1)-E(k)$, we get
$$E(k+1)=E(k)+2-q-q^{k+1}.$$
Substituting the appropriate expression for $E(k),\,E(k-1)$,\ldots, $E(1)=1$, we get
$\displaystyle E(k)=(2k-1)-(k-2)q-\frac{q-q^{k+1}}{1-q}=2k-(k-2)q-\frac{1-q^{k+1}}{1-q}.$
\end{proof}

\section{Proof of Result \ref{res:2}}
\label{A:b}
\begin{proof}
Denote $\displaystyle f(k)=E_{S}\left(k, p\right),\,\,\, \dot{f}(k)=\frac{\partial f(k)}{\partial k},\,\,\,\, \ddot{f}(k)=\frac{\partial^2 f(k)}{\partial k^2}.$
Recall (see \eqref{eq:S}) that $\displaystyle f(k)=2-q-\frac{1-q^{k+1}-2q+2q^2}{k(1-q)}$. We have $\displaystyle f(1)=1$,\,\,$\lim_{k\uparrow \infty}f(k)=2-q$,\,\
$\displaystyle \dot{f}(k)=\frac{1}{1-q}\left[\frac{1}{k^2}(1-q^{k+1}-2q+2q^2)+\frac{1}{k}q^{k+1}ln(q)\right]$. So, $\displaystyle \dot{f}(1)<0$ for
$q\in \left(\frac{\sqrt{5}-1}{2},\,1\right)$ and, therefore, the function $f(k)$ (as a function of continuous variable $k\geq 1$) has a minimum in support $k\geq 1$.
Further, $\ddot{f}(k)=\frac{1}{1-q}\left[-\frac{2(1-q)}{k}\dot{f}(k)+\frac{1}{k}q^{k+1}(\ln(q))^2\right]$ and if $\dot{f}(l)=0$, then $\ddot{f}(l)>0$, shows that $l$ is the unique minimum.
\end{proof}

\section{Proof of Result \ref{res:d}}
\label{A:d}
\begin{proof}
$N=sk^{*}_{S}(p)$.
$\displaystyle \frac{h_{S}\left(k^{*}_{S}(p)\right)}{k^{*}_{S}(p)}=E_{S}\left(k^{*}_{S}(p),\, p\right)\leq E_{S}\left(k,\, p\right)=\frac{h_{S}(k)}{k}$ for any $k=1,2,\ldots$ implies
$ \displaystyle s h_{S}\left(k^{*}_{S}(p)\right)=\frac{\sum_{i=1}^{J}m_i}{k^{*}_{S}(p)} h_{S}\left(k^{*}_{S}(p)\right)
\leq \sum_{i=1}^{J}h_{S}\left(m_i,\, p\right)$
 for any partition $\displaystyle \left\{m_1,\ldots,m_J\right\}$ with $\displaystyle \sum_{i=1}^{J}m_i=N,\,\,\,J\in\left\{1,\ldots,N\right\}$, which completes the proof.
\end{proof}

\section{Proof of Result \ref{res:LS}}
\label{A:c}
\begin{proof}
It is easy to verify that for all $p\in (0,1)$ the second derivative of $h_{S}(x)=x E_{S}\left(x, p\right)$ with respect to $x$ is positive and, therefore, the function
$h_{S}(x)$ is convex with respect to $x$. We start with some partition
$\displaystyle \left\{m_1,\ldots,m_J\right\}$. Convexity of $h_{S}(x)$ implies that for any $m_j-m_i\geq 2$, $h_{S}(m_j-1)+h_{S}(m_i+1)\leq h_{S}(m_j)+h_{S}(m_i)$.
Applying this $+1,-1$ improvement for any $i,j$ with $m_j-m_i\geq 2$, we obtain a
better (with respect to expected number of tests) partition $\left\{m^{'}_1,\ldots,m^{'}_J\right\}$  with $\displaystyle |m^{'}_j-m^{'}_i|\leq 1$ for all $i,j$.
\end{proof}

\section{Proof of Result \ref{res:Main}}
\label{A:cc}
The proof for Procedure $S$ is exactly the same as a proof for Procedures $D^{\prime}$ and $M$ in \cite{G1985} (p. 389) and is based on the fact that the function $\displaystyle f(x)=E_{S}\left(x, p\right)$ has a unique minimum for $x\geq 1$ as was shown in the proof of Result \ref{res:2} in Appendix \ref{A:b}.

\bigskip
\begin{center}
{\large\bf SUPPLEMENTARY MATERIAL}
\end{center}

\begin{description}
\item[Web Appendix F] Implementation of Result 5
\item[Web Appendix G] Ungar Construction
\item[Web Appendix H] Connection of group testing and coding theory
\item[Web Appendix I] Development of the optimal nested procedure
\item[Web Appendix J] Matlab code for the optimum nested procedure

\end{description}

{}


\begin{thebibliography}{}


\bibitem[\protect\citeauthoryear{Bar-Lev \it{et~al.}}{1990}]{BP1990}
Bar-Lev, S.~K., Boneh, A., Perry, D. (1990).
\newblock Incomplete identification models for group-testable items.
\newblock {\emph Nav. Res. Logist.} {\textbf 37,} 647--659.


\bibitem[\protect\citeauthoryear{Bar-Lev \it{et~al.}}{2005}]{B2005}
Bar-Lev, S.~K., Stadje, A., van der Duyn Schouten, F.~A. (2005).
\newblock Multinomial group testing models with incomplete identification.
\newblock {\emph J. Stat. Plan. Inf.} {\textbf 135,} 384--401.

\bibitem[\protect\citeauthoryear{Bar-Lev \it{et~al.}}{2017}]{B2016}
Bar-Lev, S.~K., Boxma, O., Kleiner, I., Perry, D. (2017).
\newblock Recycled incomplete identification procedures for blood screening.
\newblock {\emph Eur. J. Oper. Res.}, {\textbf 259,} 330--343.


\bibitem[\protect\citeauthoryear{Bellman}{1957}]{B1957}
Bellman, R. (1957).
\newblock Dynamic Programming. {\it Princeton University Press}.


\bibitem[\protect\citeauthoryear{Bilder \it{et~al.}}{2010}]{BT2010}
Bilder, C.~R., Tebbs, J.~M., Chen, P. (2010).
\newblock Informative retesting.
\newblock {\emph J. Am. Stat. Assoc.} {\textbf 105,} 942--955.

\bibitem[\protect\citeauthoryear{Brady and Greighton}{2000}]{BG2000}
Brady, P., Greighton, T. (2000).
\newblock Searching for periodic sources with LIGO. II: Hierarchical searches.
\newblock {\emph Phys. Rev. D} {\textbf 61,} 082001.



\bibitem[\protect\citeauthoryear{Cao and Sun}{2016}]{CS2016}
Cao, C., Sun, X. (2016).
\newblock Combinatorial pooled sequencing: experiment design and decoding.
\newblock
{\emph Quantitative Biology} {\textbf 4,} 36--46.




\bibitem[\protect\citeauthoryear{Delaigle and Hall}{2012}]{DH2012}
Delaigle, A., Hall, P. (2012).
\newblock Nonparametric regression with homogeneous group testing data.
\newblock {\it Ann. Statist.} {\textbf 40,} 131--158.



\bibitem[\protect\citeauthoryear{Dood \it{et~al.}}{2002}]{D2002}
\newblock
Dodd, R.Y., Notari IV, E.P., Stramer, S.L. (2002).
\newblock
Current prevalence and incidence of infectious disease markers and estimated window-period risk in the American Red Cross blood donor population.
\newblock
{\it Transfusion} {\textbf 42,} 975--979.


\bibitem[\protect\citeauthoryear{Dorfman}{1943}]{Dorfman1943}
Dorfman, R. (1943).
\newblock The detection of defective members of large populations.
\newblock {\emph The Annals of Mathematical Statistics } {\textbf 14,} 436--440.


\bibitem[\protect\citeauthoryear{Du and Hwang}{1999}]{Dh1999}
Du, D., Hwang, F. K. (1999).
\newblock Combinatorial Group Testing and its Applications. {\it World
Scientific, Singapore}.

\bibitem[\protect\citeauthoryear{Du and Hwang}{2006}]{Dh2006}
Du, D., Hwang, F. K. (2006).
\newblock Pooling Design and Nonadaptive Group Testing: Important Tools for DNA Sequencing. {\it World
Scientific, Singapore}.


\bibitem[\protect\citeauthoryear{Du and Ko}{1987}]{DK1987}
Du, D.Z., and Ko, K.I. (1987).
\newblock Some completeness results on decision trees and group testing.
\newblock {\emph SIAM. J. on Algebraic and Discrete Methods} {\textbf 8,} 762--777.




\bibitem[\protect\citeauthoryear{Feller}{1950}]{F1950} Feller, W. (1950).
\newblock
{An introduction to probability theory and its application}.
{\it New York: John Wiley \& Sons}.


\bibitem[\protect\citeauthoryear{France \it{et~al.}}{2015}]{F2015}
\newblock
France, B., Bell, W., Chang, E., Scholten, T. (2015).
\newblock
Composite sampling approaches for Bacillus anthracis surrogate extracted from soil.
\newblock
{\it PLoS One} {\textbf 10(12),} 1--18.





\bibitem[\protect\citeauthoryear{Garey and Johnson}{1979}]{GJ1979}
Garey, M.R, and Johnson, D. S. (1979).
\newblock Computers and Intractability. A Guide to the Theory of NP-Completeness.
{\it  W. H. Freeman and company}.


\bibitem[\protect\citeauthoryear{Gastwirth and Johnson}{1994}]{G1994}
Gastwirth, J. and Johnson, W. (1994).
\newblock
Screening with cost effective
quality control: Potential applications to HIV and drug testing.
\newblock
{\emph J. Amer. Statist. Assoc.} {\textbf 89,} 972--981.


\bibitem[\protect\citeauthoryear{Gilstein}{1985}]{G1985}
Gilstein, C. Z. (1985).
\newblock
Optimal partitions of finite populations for Dorfman-type group testing.
\newblock
{\it J. Stat. Plan. Inf.} {\textbf 12,} 385--394.



\bibitem[\protect\citeauthoryear{Gr{\ae}sb{\o}ll \it{et~al.} }{2016}]{G2016}
Gr{\ae}sb{\o}ll, K., Andresen, L-O., Halasa, T., Toft, N.(2016).
\newblock
Opportunities and challenges when pooling milk samples using ELISA.
\newblock
{\it Preventive Veterinary Medicine} {\textbf 139} Part B, 93--98.


\bibitem[\protect\citeauthoryear{Haber \it{et~al.}}{2017}]{Ha2016}
\newblock
Haber, G., Malinovsky Y., and Albert, P.~S.(2017).
\newblock
Sequential estimation in the group testing problem.
\newblock
{\it Sequential Analysis}. In press.


\bibitem[\protect\citeauthoryear{Hill \it{et~al.}}{2016}]{Hill2016}
\newblock
Hill, J.~A., HallSedlak, R., Magaret, A., Huang, M.~L., Zerr, D.~M., Jeromeb, K.~R., Boeckh, M.(2016).
\newblock
Efficient identification of inherited chromosomally integrated humanherpesvirus 6 using specimen pooling.
\newblock
{\it Journal of Clinical Virology } {\textbf 77,} 71--76.

\bibitem[\protect\citeauthoryear{Hu and Tucker}{1971}]{HT1971}
Hu, T. C., and Tucker, A.C. (1971).
\newblock Optimum computer search tree.
\newblock {\it SIAM Journal on Applied Mathematics } {\textbf 21,} 514--532.


\bibitem[\protect\citeauthoryear{Huffman}{1952}]{H1952}
Huffman, D. A. (1952).
\newblock A Method for the Construction of Minimum-Redundancy Codes.
\newblock {\emph Proceedings of the I.R.E. } {\textbf 40,} 1098--1101.



\bibitem[\protect\citeauthoryear{Hwang}{1976}]{H1976}
Hwang, F. K. (1976).
\newblock
An optimal nested procedure in binomial group testing.
\newblock
{\it Biometrics} {\textbf 32,} 939--943.













\bibitem[\protect\citeauthoryear{Kim {\it et~al.}}{2007}]{Kim2007} Kim, H. Y., Hudgens, M. G., Dreyfuss, J. M., Westreich, D. J., and Pilcher, C. D. (2007).
Comparison of group testing algorithms for case indentification in the presence of test error.
{\it Biometrics} {\bf 63,}  1152--1162.

\bibitem[\protect\citeauthoryear{Kumar and Sobel}{1971}]{KS1971}
Kumar, S., and Sobel, M.  (1971).
\newblock
Finding a single defective in binomial group-testing.
\newblock
{\it J. Amer. Statist. Assoc.} {\textbf 66,} 824--828.


\bibitem[\protect\citeauthoryear{Laarhoven}{2013}]{L12013}
Laarhoven, T. (2013).
\newblock
Efficient probabilistic group testing based on traitor tracing.
\newblock
{\it 51st Annual Allerton Conference on Communication, Control and Computing, At Monticello IL, USA}.


\bibitem[\protect\citeauthoryear{Lee and Sobel}{1972}]{LS1972}
Lee, J.K., and Sobel, M. (1972).
\newblock Dorfman and $R_1$-type procedures for a generalized group testing problem.
\newblock {\it Mathematical Biosciences} {\textbf 15,} 317--340.

\bibitem[\protect\citeauthoryear{Liu \it{et~al.}}{2012}]{Liu2012}
Liu, A., Liu, C.~L., Zhang, Z., Albert, P.~S. (2012).
\newblock Optimality of group testing in the presence of misclassification.
\newblock {\it Biometrika} {\textbf 99,} 245--251.





\bibitem[\protect\citeauthoryear{Malinovsky and Albert}{2015}]{MA2015} Malinovsky, Y., Albert, P. S. (2015).
A note on the minimax solution for the two-stage group testing problem.
{\it The American Statistician} {\textbf 69,} 45--52.

\bibitem[\protect\citeauthoryear{Malinovsky \it{et~al.}}{2016}]{MAR2016} Malinovsky, Y., Albert, P. S., Roy, A. (2016).
Reader reaction: A note on the evaluation of group testing algorithms in the presence of misclassification.
{\it  Biometrics} {\textbf 72,} 299--302.

\bibitem[\protect\citeauthoryear{McMahan \it{et~al.}}{2012}]{M2012} McMahan, C. S., Tebbs, J. M., and  Bilder, C. R. (2012).
Informative Dorfman screening
{\it Biometrics} {\bf 68,}  287--296.

\bibitem[\protect\citeauthoryear{Meinshausen \it{et~al.}}{2009}]{MBR2009}
Meinshausen, N., Bickel, P., Rice, J. (2009).
\newblock Efficient blind search: optimal power of detection under computational cost constraints.
\newblock {\it The Annals of Applied Statistics} {\textbf 3,} 38--60.



\bibitem[\protect\citeauthoryear{Moon and Sobel}{1977}]{MS1977} Moon, J. W., Sobel, M. (1977).
Enumerating a class of nested group testing procedures.
{\it Journal of combinatorial theory, series B} {\textbf 23,} 184--188.

\bibitem[\protect\citeauthoryear{Pfeifer and Enis}{1978}]{PE1978}
Pfeifer, C. G., Enis, P. (1978).
\newblock
Dorfman-type group testing for a modified binomial model.
\newblock
{\it J. Amer. Statist. Assoc.} {\textbf 73,} 588--592.






\bibitem[\protect\citeauthoryear{Samuels}{1978}]{S1978}
Samuels, S. M. (1978).
\newblock The exact solution to the two-stage group-testing problem.
\newblock {\it Technometrics} {\textbf 20,} 497--500.

\bibitem[\protect\citeauthoryear{Sobel and Groll}{1959}]{SG1959}
Sobel, M., Groll, P. A. (1959).
\newblock Group testing to eliminate efficiently all defectives in a binomial sample.
\newblock {\it Bell System Tech. J.} {\textbf 38,} 1179--1252.

\bibitem[\protect\citeauthoryear{Sobel}{1960}]{S1960}
Sobel, M. (1960).
\newblock Group testing to classify efficiently all defectives in a binomial sample.
\newblock {\it Information and Decision Processes (R. E. Machol, ed.; McGraw-Hill, New York),} pp. 127-161.



\bibitem[\protect\citeauthoryear{Sobel}{1967}]{S1967}
Sobel, M. (1967).
\newblock Optimal group testing.
\newblock {\it Proc. Colloq. on Information Theory, Bolyai Math. Society, Debrecen, Hungary}.



\bibitem[\protect\citeauthoryear{Sterrett}{1957}]{S1957}
Sterrett, A. (1957).
\newblock On the detection of defective members of large populations.
\newblock {\emph The Annals of Mathematical Statistics } {\textbf 28,} 1033--1036.


\bibitem[\protect\citeauthoryear{Stramer \it{et~al.}}{2011}]{S2011}
Stramer, S.~L., Wend, U., Candotti, D., Foster, G.~A.,
Hollinger, F.~B., Dodd, R.~Y., Allain, J.~P., Gerlich, W. (2011).
\newblock Nucleic acid testing to detect HBV infection in blood donors.
\newblock {\emph The New England Journal of Medicine} {\textbf 364,} 236--247.



\bibitem[\protect\citeauthoryear{Tebbs \it{et~al.} }{2013}]{T2013}
Tebbs, J., McMahan, C., and Bilder, C. (2013).
\newblock Two-stage hierarchical group testing for multiple infections with application to the Infertility Prevention Project.
\newblock {\emph Biometrics} {\textbf 69,} 1064--1073.









\bibitem[\protect\citeauthoryear{Thompson}{1962}]{To1962}
Thompson, K.H. (1962).
\newblock
Estimation of the proportion of vectors in a natural population of insects.
\newblock
{\it Biometrics} {\bf 18}, 568--578.

\bibitem[\protect\citeauthoryear{Tu \it{et~al.}}{1995}]{Tu1995}
Tu, X.~M., Litvak, E., and Pagano, M.(1995).
\newblock
On the information and accuracy of pooled testing in estimating
prevalence of a rare disease: Application to HIV screening.
\newblock
{\it Biometrika} {\textbf 82,} 287--297.

\bibitem[\protect\citeauthoryear{Ungar}{1960}]{U1960}
Ungar, P. (1960). Cutoff points in group testing.
{\it Comm. Pure Appl. Math.} {\bf 13,} 49--54.







\bibitem[\protect\citeauthoryear{Warasi \it{et~al.}}{2016}]{Wa2016}
Warasi, M., Tebbs, J., McMahan, C., and Bilder, C. (2016).
\newblock
Estimating the prevalence of multiple diseases from two-stage hierarchical pooling.
\newblock
{\it Stat. Med.} {\bf 31,} 185--191.


\bibitem[\protect\citeauthoryear{Wolf}{1985}]{W1985}
Wolf J. K. (1985). Born again group testing: multiaccess comunications.
{\it IEEE Transactions on Information Theory} {\bf 31,} 185--191.



\bibitem[\protect\citeauthoryear{Yao and Hwang}{1988}]{YH1988}
Yao, Y. C., Hwang, F. K. (1988).
\newblock
A fundamental monotonicity in group testing.
\newblock
{\it SIAM J. Discrete Math.} {\textbf 1,} 256--259.


\bibitem[\protect\citeauthoryear{Yao and Hwang}{1990}]{YH1990}
Yao, Y. C., Hwang, F. K. (1990).
\newblock
On optimal nested group testing algorithms.
\newblock
{\it J. Stat. Plan. Inf.} {\textbf 24,} 167--175.




\bibitem[\protect\citeauthoryear{Zaman and Pippenger}{2016}]{ZP2016}
Zaman, N., and Pippenger, N. (2016).
\newblock
Asymptotic analysis of optimal nested group-testing procedures.
\newblock
{\it Prob. Eng. Inform. Sci.} {\textbf 30,} 547–-552.

\bibitem[\protect\citeauthoryear{Zhu \it{et~al.}}{2001}]{Z2001}
Zhu, L., Hughes-Oliver, J.~M, Young, S.~S. (2001).
Statistical decoding of potent pools based on chemical structure.
{\it Biometrics} {\textbf 57,} 922--930.


\end{thebibliography}
\end{document}


\def\spacingset#1{\renewcommand{\baselinestretch}%
{#1}\small\normalsize} \spacingset{1}


\if0\blind
{
  \title{\bf Supplementary Materials\\
  Revisiting nested group testing procedures: new results, comparisons and robustness}
  \author{Yaakov Malinovsky\thanks{Corresponding author}\\
    Department of Mathematics and Statistics\\ University of Maryland, Baltimore County, Baltimore, MD 21250, USA\\
    and \\
    Paul S. Albert\thanks{
    The work was supported by the National Cancer Institute Intramural Program.}\hspace{.2cm}\\
    Biostatistics Branch, Division of Cancer Epidemiology and Genetics\\National Cancer Institute, Rockville, MD 20850, USA}

  \maketitle
} \fi

\if1\blind
{
  \bigskip
  \bigskip
  \bigskip
  \begin{center}
    {\LARGE\bf Title}
\end{center}
  \medskip
} \fi

\newpage
\spacingset{1.45} 

\section*{Web Appendix}
\appendix

\setcounter{section}{5}
\section{Implementation of Result 5}
\label{A:e}
\begin{remark}
\begin{itemize}
\item[(i)]
From Result 5, it follows that under a partition into $s$ groups of sizes $\displaystyle l=\left\lfloor\frac{N}{s}\right\rfloor$  and $\displaystyle l+1$ (or alternatively $\displaystyle l+1=a+\left\lfloor\frac{\theta}{s}\right\rfloor$), we have the following relationship:
$$\displaystyle l g_{l}+(l+1)g_{l+1}=N,\,\,\,g_l+g_{l+1}=s,$$
where $g_l$ is the number of groups of size $l$. The trivial solution of these equations shows that we have
$g_l=s(l+1)-N$ and $g_{l+1}=N-sl$.
\item[(ii)]
From Result 5, it follows that under a partition into $s+1$ groups of sizes $\displaystyle w=\left\lfloor\frac{N}{s+1}\right\rfloor$  and $\displaystyle w+1$ (or alternatively $\displaystyle w+1=a+\left\lfloor\frac{\theta}{s}\right\rfloor$) we have the following relations:
$$\displaystyle w g_{w}+(w+1)g_{w+1}=N,\,\,\,g_w+g_{w+1}=s+1,$$
The trivial solution of these equations shows that we have
$g_w=(s+1)(w+1)-N$ and $g_{w+1}=N-w(s+1)$.
\end{itemize}
\end{remark}

\section{Ungar Construction}
\label{se:U}

We recall that Ungar's (1960) fundamental result examines the general problem of when group testing is preferable to individual testing.\\
{\it Result from \cite{U1960}:
The individual testing is optimal if and only if $\displaystyle p\geq \frac{3-\sqrt{5}}{2}.$
}\\

We need to show Ungar's (1960) construction for $N=2$ to prove future results.
The tree below presents a reasonable group testing algorithm for $N=2$ as it was presented by \cite{U1960} (the left branch of the tree represent the negative test result, and the right branch represents the positive test result):

\Tree [.{test 1 and 2} [.{stop (T=1)} ] [.{test 1} [.{stop (T=2)} ] [.{test 2 (T=3)}  ] ] ]

The expected number of tests per person is $\displaystyle \frac{1}{2}\left(3-q-q^2\right).$  Comparing it with 1, we conclude that this algorithm is better than individual testing when $\displaystyle p<\frac{3-\sqrt{5}}{2}$ or $\displaystyle q >\frac{\sqrt{5}-1}{2}$.

\section{Connection of group testing and coding theory}
\label{A:cod}

The connection of group testing with noiseless-coding theory was presented in the group testing literature by \cite{SG1959} and further investigated
in \cite{S1960, S1967},  \cite{KS1971}, \cite{H1974, H1976}, \cite{GK1976}, \cite{W1985} and \cite{YH1990}. For a comprehensive discussion, see \cite{K1973}.

To demonstrate the connection, first consider the case when $N=2$. There are $4$ possible outcomes that correspond to the total number of tests $T=1,2,3$. We use the code $-$ if a person is free from decease and
$+$ otherwise.
Therefore, there are $M=2^2=4$ possible states of nature:
$$--\,\,\,\,\text{with probability}\,\,\,\,\,\,\,\,\,\,\,\,\,\,\,\,\,\,\,\, q^2, $$
$$-+\,\,\,\,\text{with probability}\,\,\,\,\, q(1-q), $$
$$+-\,\,\,\,\text{with probability}\,\,\,\,\, q(1-q), $$
$$++\,\,\,\,\text{with probability}\,\,\,\,\, (1-q)^2.$$

We use a binary (in which only the binary digits 0 and 1 are employed) {\it prefix code} for these 4 possible states of nature.
The code is called a prefix code if no code word is the prefix (contained fully as the beginning) of
any other code word. We want to code these 4 states with the binary prefix code with lengths  $l_1,l_2,l_3,l_4$ such that the expected length
$$p_1l_1+p_2l_2+p_3l_3+p_4l_4,$$
will be minimal, where $\displaystyle \left\{p_1,p_2,p_3,p_4\right\}=\left\{q^2,q(1-q),q(1-q),(1-q)^2\right\}$.  The solution of this problem is due to \cite{H1952} and the simple explanation here is due to \cite{R1984}. Without loss of generality, assume $\displaystyle p_1\geq p_2 \geq p_3 \geq p_4$.
Denote $L_4$ as the minimal expected length.
For $M=2$ with $\displaystyle p_1\geq p_2$, the problem is trivial. For $M=3,$ with $\displaystyle p_1\geq p_2 \geq p_3$, it has to be decided which of these probabilities should be assigned
to the point which is one unit away from the root. It is clear that
it must be the largest of the three probabilities (i.e., $p_1$), and that the two smaller ones ($p_2$ and $p_3$)
should be assigned to the point which is 2 units away. Below is the assignment and the optimal prefix code as a tree presentation for $M=3$:

\Tree [.{root} [.{0} ] [.{} [.{10} ] [.{11}  ] ] ]

As it was noticed, to the largest $p_1$ is assigned code $0$, and to the two others, $p_2$, and $p_3$, are assigned codes $10$ and $11$.
Now it is clear how to proceed with $M=4$ with $p_1\geq p_2 \geq p_3 \geq p_4$. Denote $p_{34}=p_3+p_4$. In the previous case, $M=3$ and we know how to construct a tree. On this tree, we will branch
out two new branches from the node with $p_{34}$ and put the numbers $p_3$ and
$p_4$ at the two terminal nodes.

Now we can apply this optimal construction to the group testing with $N=2$ ($M=4$) For $q\geq 1/4,$ we have
$p_1=q^2 \geq  p_2=q(1-q) \geq  p_3=q(1-q) \geq p_4=(1-q)^2,$ and for $q\geq (\sqrt{5}-1)/2$ we have $p_1\geq p_{34}=p_3+p_4=1-q\geq p_2.$
\begin{center}
\tiny
\Tree [.{prob. 1} [.{with prob. $q^2$ (code 0)} ] [.{prob. 1-q^2} [.{with prob. q(1-q) (code 10)} ] [.{prob. 1-q} [.{prob. q(1-q) (code 110)}  ][.{prob. (1-q)^2 (code 111)}  ] ] ] ]
\end{center}
Therefore, $L_4=3-q-q^2$, and this is equal to the expected number of tests in the Ungar construction with $N=2$.

\section{Development of the optimal nested procedure}
\label{A:n}
The definition of nested procedure $R_1$ was presented in Section 3 of the article. We need the following \citep{S1960} result and lemma. We will cite them verbatim.\\
For a defective set of size $m\geq 1$, denote the number of defective units $Y$ and denote by $Z$ the number of defectives presents in the subset of size $x$ (with with $1\leq x \leq m-1$) randomly chosen from the defective set, then
\begin{equation}
\label{eq:SS}
P\left(Z=0\,|\,Y\geq 1\right)=\frac{q^{x}(1-q^{m-x})}{1-q^m}.
\end{equation}

{\it Lemma 1. Given a defective set of size $m\geq 2$ and given that a proper subset of size $x$ with $1\leq x \leq m-1$ also proves to contain at least one defective, then the posteriori distribution associated with $m-x$ remaining units is precisely that $m-x$ independent binomial chance variables with common probability $q$ of being good.}

Let $G(m,n)$ denote the minimum expected number of tests needed to classify all $m$ units in defective set and the remaining $n-m$ units in a binomial set (G-situation). Define $H_1(n)=G(0,n)$ (H-situation). Then, from \eqref{eq:SS} and Lemma 1 above, we have the following recursion (dynamic programming) equation which will lead as to the goal $H_{1}(N)$:
\begin{align}
\label{eq:dp1}
&
H_1(0)=0,\,H_1(1)=1,\nonumber
\\
&\displaystyle
H_1(n)=1+\min_{1\leq x \leq n}\left\{q^{x}H_1(n-x)+(1-q^{x})G(x,n)\right\},\,\,\,n=2,\ldots,N,\\\nonumber
&
G(1,n)=H_1(n-1)\,\,\,\,n=1,2,\ldots,N,\\\nonumber
&
\displaystyle
G(m,n)=1+\min_{1\leq x \leq m-1}\left\{\frac{q^x-q^m}{1-q^m}G(m-x,n-x)+\frac{1-q^{x}}{1-q^m}G(x,n)\right\},\,2\leq m \leq n=2,\ldots,N.
\end{align}

It can be verified that the complexity of the calculation using above DP algorithm \eqref{eq:dp1} is $O(N^3)$.

\cite{S1960} reformulated the problem and found a solution with complexity $O(N^2)$ in the following way.
The iteration equations \eqref{eq:dp1} for $n=2,\dots, N$ will always lead to the ``break down" defective set of size $m\geq 2$. In particular, if the unit $i$ is the first defective units
in the defective set that we found, then we come to the $H$-situation with $n-i$ units (follows from Lemma 1 above). This observation allows us to proceed in the following way.
Denote $F_1(m)$ as the minimum expected number of tests required to ``break down" a defective set of size size $m\geq 2$ and for the first time reach $H$-situation when $q$ is given. Then $F_1(m)$ does not depend on $n$ and we can write
\begin{align}
\label{eq:dp2}
&
G(m,n)=F_1(m,n)+\sum_{i=1}^{m}\frac{q^{i-1}p}{1-q^m}H(n-i),\,\,\,2\leq m \leq n=2,\ldots,N.
\end{align}

Denote $$\displaystyle F^{*}_{1}(m)=\frac{1-q^m}{1-q}F_{1}(m),\,\,G^{*}(m,n)=\frac{1-q^m}{1-q}G(m,n).$$
Then combining \eqref{eq:dp1} with \eqref{eq:dp2} leads to the improved dynamic programming algorithm (below) with complexity $O(N^2)$:

\begin{align}
\label{eq:dpi}
&
H_1(0)=0,\,H_1(1)=1,\nonumber
\\
&\displaystyle
H_1(n)=1+\min_{1\leq x \leq n}\left\{q^{x}H_1(n-x)+(1-q)\left[F_1^{*}(x)+\sum_{i=1}^{x}q^{i-1}H_1(n-i)\right]\right\},\,\,\,n=2,\ldots,N,\\ \nonumber
&
F_1^{*}(1)=0,\\
&
F_1^{*}(m)=\frac{1-q^m}{1-q}+\min_{1\leq x \leq m-1}\left\{q^x F_1^{*}(m-x)+F_1^{*}(x)\right\},\,\,\,\, m=2,\ldots,N. \nonumber
\end{align}

\bigskip
We demonstrate the construction of an optimum nested algorithm with the following example.\\
{\it Example}: $p=0.05,\,\, N=13$\\
\begin{table}[H]
\begin{center}
\begin{tabular}{|l|l|l|l|l|l|l|l|l|l|l|l|l|}
  \hline
  n& 13 & 12 & 11 & 10 & 9 & 8 & 7 & 6 & 5 & 4 & 3 & 2\\
  $x_H$ & 13 & 12 & 11 & 10 & 9 & 8 & 7 & 6 & 5 & 4 & 3 & 2\\
  $x_G$ & 5 & 4 & 4 & 4 & 4 & 4 & 3 & 2 & 2 & 2 & 2 & 1\\
\hline
\end{tabular}
\end{center}
\end{table}
In this table, $n$ is a size of the set $C$ that still is not yet classified, $x_H$ is the size of the subset that we have to check given that the set $C$ is the binomial set (H-situation) and $x_G$ is the size of the subset that we have to check given that set $C$ is the defective set (G-situation).

\section{Matlab code for the optimum nested procedure}
\begin{enumerate}
\item[(i)]
{\bf Matlab function ``Fstar"}\\
\#function  [Fs Loc]=Fstar(p,n)\\
\#$q=1-p$;\\
\#$Fs=zeros(n,1)$; $Fs(1,1)=0$; $Loc=[]$;\\
\#for $m=2:1:n$\\
    \#$f=floor(m/2)$; $l=zeros(f,1)$;\\
    \#for $x=1:1:f$\\
        \#$l(x)=(q^x)*Fs(m-x)+Fs(x)$;\\
    \#end\\
    \#$Fs(m)=(1-q^m)/(1-q)+min(l)$;\\
    \#\%Location of min loc\\
    \#$ll=min(l)$; $loc=find(l==ll)$; $Loc=[Loc; m\,\,\,\,\,\, loc ]$;\\
\#end  end
\item[(ii)]
{\bf Matlab function ``H1"}\\
\#function $[H_1\,\, D]= H1(p,n)$\\
\#$q=1-p$; $[Fs\,\, Loc]=Fstar(p,n)$;\\
\#$H_1=zeros(n,1)$; $H_1(1,1)=1$; $D=[]$;\\
\#for $N=2:1:n$\\
   \#$l=zeros(N,1)$;\\
    \#$qN=((fliplr(eye(N-1)))*(q.^[0:1:N-2]')).*H_1(1:N-1,1)$;\\
    \#$S=[]$;\\
    \#for $x=1:1:N-1$\\
        \#$l(x)=(q^x)*H_1(N-x)+(1-q)*(Fs(x)+sum(qN(((N-1)-(x-1)):(N-1),1)))$;\\
        \#$S=[S; 1+l(x)\,\, N\,\, x ]$;\\
    \#end\\
    \#$l(N)=(1-q)*(Fs(N)+sum(qN))$;\\
    \#$S=[S; 1+l(N)\,\, N\,\,\, N ]$; $SS=S(S(:,1)==min(S(:,1)),:)$;\\
    \#$D=[D;SS]$; $H_1(N)=1+min(l)$; end\\
\#$D=D(:,2:end)$;
\end{enumerate}

{}